# An FBAR Circulator

Mustafa Mert Torunbalci, Trevor J. Odelberg, *Student Member, IEEE*, Suresh Sridaran, *Member, IEEE*, Richard C. Ruby, *Fellow, IEEE*, and Sunil A. Bhave, *Senior Member, IEEE*

*Abstract*—This letter presents the experimental demonstration of a film bulk acoustic resonator (FBAR) circulator at 2.5 GHz. The circulator is based on spatiotemporal modulation of the series resonant frequency of FBARs using varactors and exhibits a large isolation of 76 dB at 2.5 GHz. The FBAR chip (0.25 mm$^2$) consists of three identical FBARs connected in wye configuration. FBAR's quality factor ($Q$) of 1250 and the piezoelectric coupling coefficient $k_t^2$ of 3% relax the modulation requirements, achieving nonreciprocity with small modulation to RF ratio better than 1:800 (3 MHz: 2.5 GHz).

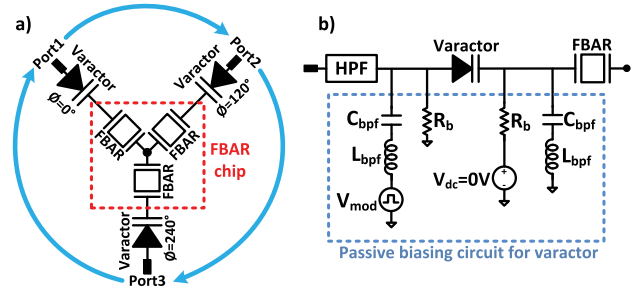

Fig. 1. Schematic view (a) FBAR chip at the center of wye topology connects to three varactors to form the circulator. (b) The varactor biasing network identical to [2]. The bpf ensures that modulation does not leak into the FBAR chip and the high-pass filter (hpf) allows RF signals to and from the circulator without the modulation leaking into the network analyzer.

TABLE I
COMPONENT LIST AND VALUES FOR FBAR CIRCULATOR

| Component | Name | Value |
|---|---|---|
| $Varactor$ | MACOM-MA46H120 | $C_{var}$=1 pF @ 0 V DC |
| $L_{bpf}$ | Coilcraft 03CS29N | 29 nH |
| $C_{bpf}$ | ATC 530L | 100 nF |
| $R_b$ | ERA-6AEB105V | 1 MΩ |
| $HPF$ | Minicircuit VHF-1200+ | IL=2 dB from 1.2-4.6 GHz |

## I. INTRODUCTION

CIRCULATORS are nonreciprocal three-terminal devices that transmit electromagnetic signals entering any port to the next port, only in one direction. These devices play a crucial role in communication systems: 1) protecting the laser from reflected signals as an isolator in photonic systems and 2) enabling the transmitter and the receiver to share the same frequency band in full-duplex wireless systems. The most common approach and commercially available technology for breaking reciprocity to realize a circulator is to use an external magnetic field in a ferromagnetic medium. However, this approach requires bulky magnetics that are expensive and not CMOS compatible. Recently, alternative approaches have been demonstrated in [1] and [2], where the three identical coupled $LC$ resonators are modulated via spatiotemporal modulation using ring or wye topology. These approaches provide strong nonreciprocity of 50 dB with reasonable insertion loss and bandwidth.

In this letter, we demonstrate an alternative solution to the $LC$ circulator using three identical film bulk acoustic resonators (FBARs) fabricated in wye configuration in a single packaged die. The FBARs are superior to the $LC$ lumped element resonators with a smaller footprint and higher mechanical quality factor ($Q_{FBAR}$ = 1250 at 2.5 GHz versus $Q_{LC}$ < 10 at 130 MHz) [3]. The high mechanical quality factor enables the use of relatively low modulation frequency, which will directly reduce the power consumption of the circulator. The FBARs are modulated by varactors with a phase difference of 120°, providing an electrical rotation of the RF signal with strong isolation of 76 dB. Furthermore, the FBAR circulator demonstrates the approach in [2] at a higher frequency than previously achieved using lumped element inductors and capacitors.

## II. DESIGN

The operation principle and architecture of the FBAR circulator are identical to the $LC$ wye circulator [2]. Fig. 1 presents the schematic view of the FBAR circulator, contrasting the simplicity of the FBAR chip with printed circuit board (PCB) complexity of providing the modulation signal to the varactor. Table I provides a list of components used in each branch of the circulator. The $LC$ bandpass filter (bpf) has a linear wideband characteristic which allows us to explore different modulation frequencies and amplitudes. The modulation frequency was chosen to be 3 MHz as this relates to the 3-dB bandwidth of the FBAR. The modulation signal is a 7 V$_{pp}$ square wave with 0-V dc bias that drives the varactor into forward bias with low ON impedance during the positive half cycle, and provides 0.2-pF capacitance during the negative half cycle. The modulation is not allowed to leak into the wye network by a second bpf and bleed resistor. An inline high-pass filter is placed at each RF port to prevent the high voltage modulation signal from damaging receivers in the network analyzer, at the cost of 2-dB RF insertion loss. Fig. 2 shows the effect of varactor modulation on the admittance of an FBAR. The MACOM varactor was chosen based on its tuning ratio and RF performance characteristics at gigahertz range in the datasheet.

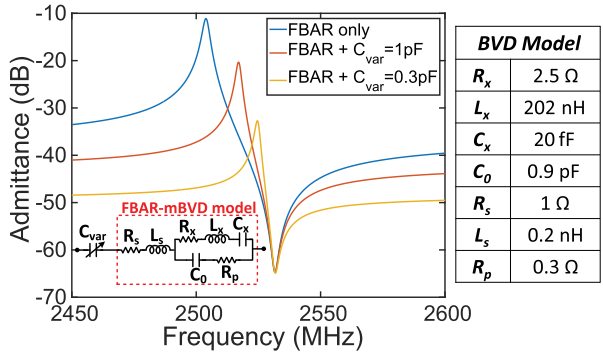

Fig. 2. Simulation of the varactor's impact on admittance of an FBAR. FBAR's BVD parameters are extracted from a fabricated resonator. Tuning the capacitance value changes the series resonance frequency at the cost of decreased admittance.

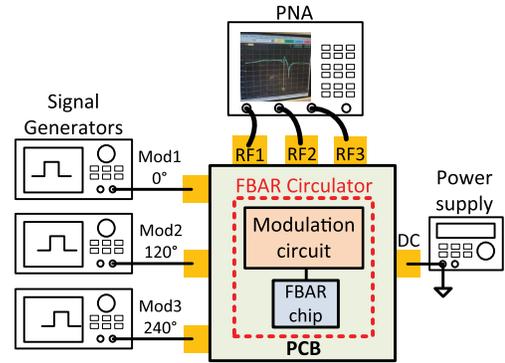

Fig. 4. Experimental setup for the characterization of FBAR circulators.

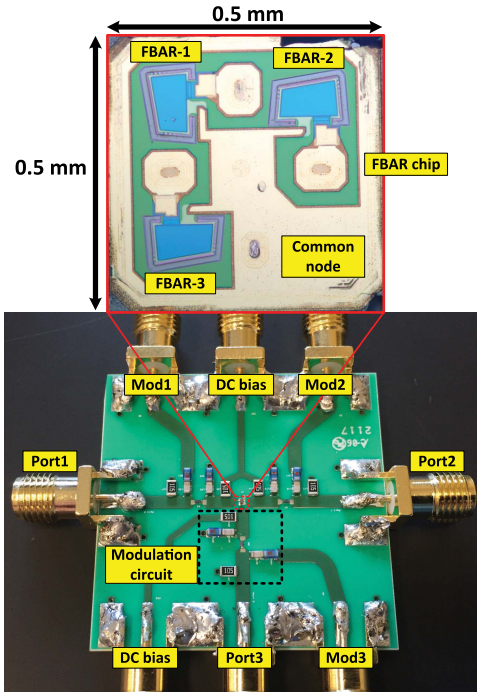

Fig. 3. Photos of the PCB where all the circulator components occupy an area of $12 \times 8$ mm$^2$. The FBAR die at the center is the smallest component of the circulator: 98% of the area is occupied by the components surrounding the varactors.

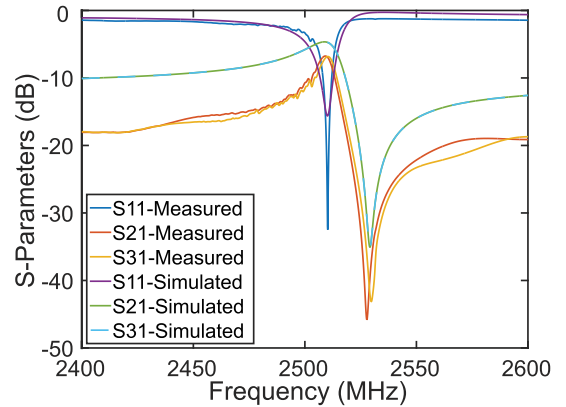

Fig. 5. Comparison of measured versus simulated $S$ parameters at port 1 when the modulation is OFF. The measured S21 has 3 dB higher loss than simulation because of PCB parasitics and from the inline high-pass filter. Measured S21 and S31 are identical confirming reciprocal behavior with no modulation.

## III. EXPERIMENTAL RESULTS

Three FBARs with series resonance frequency of 2.5 GHz were fabricated in a centroid configuration to achieve identical frequency and impedance characteristics. Fig. 3 shows the photos of the PCB where all the circulator components occupy an area of 96 mm$^2$. The FBAR die at the center is the smallest component, 98% of the area is occupied by the modulation circuits surrounding the varactors.

Fig. 4 shows a schematic view of the experimental setup used to characterize the FBAR circulators. Square wave modulation signals are generated with a phase difference of 120°, and S-parameters are measured using an Agilent four-port network analyzer. The FBAR die plus varactors have expected port to port impedance of 60 Ω based on an extracted Butterworth-Van Dyke (BVD) model. As the PCB was not optimized for high frequency using 3-D planar electromagnetic simulators, as well as the addition of the inline high-pass filter to protect the network analyzer, the measured S21 has 3 dB higher insertion loss for the FBAR chip (Fig. 5). When the modulation is OFF, S21 and S31 demonstrate identical reciprocal response. Note that the resonance frequency is shifted from FBAR's resonance frequency (Fig. 2) due to the series capacitance of the varactor. As the MACOM varactor has 13-V breakdown voltage, an initial sinusoidal modulation of 6 $V_{pp}$ and 6-V dc bias was applied to achieve the largest capacitance swing. Less than 10-dB isolation was measured with this scheme. It was discovered, however, against our original intuition that increasing the modulation voltage amplitude such that the varactor went into forward bias, and clipped and improved nonreciprocity to 12 dB. This result revealed that switchlike modulation of the varactor improved response, and the best modulation scheme was determined to be a 50% duty cycle square wave, 0-V dc bias, and 7 $V_{pp}$. As soon as the modulation signals are applied, a nonreciprocal response is observed between S21 and S31. S31 achieves an isolation of 28 dB while S21 is 12 dB with 50-Ω termination provided by the network analyzer to all three terminals (Fig. 6). The sidebands are due to the varactor nonlinearity and are spaced by 3 MHz on either side of the signal band; however, the right sideband falls in the parallel resonance of the FBAR and is greatly suppressed. The drop in S21 (modulation ON) versus S21 (modulation OFF) of 4 dB represents the RF energy that is lost to intermodulation products in this single-ended circulator topology. The measured isolation was identical at all three ports.

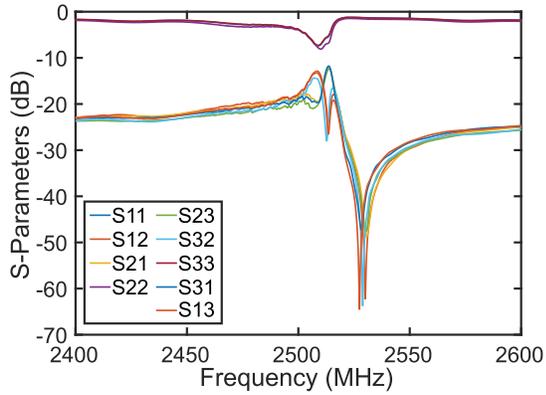

Fig. 6. When the modulation is turned ON, the measured *S* parameters at all ports exhibit forward insertion loss of 11 dB and reverse isolation of 28 dB. The measurement was performed with all three ports connected to 50 Ω impedance of the network analyzer.

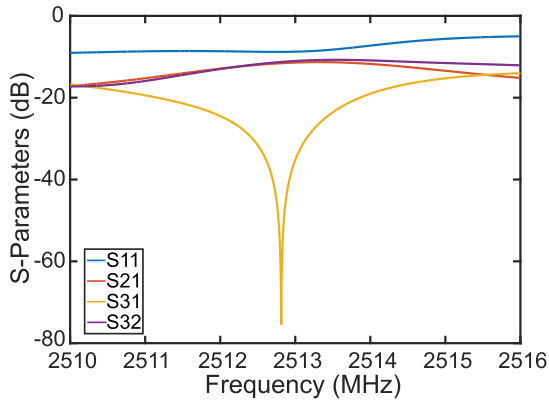

Fig. 7. Measured circulator response at port 1, after matching network of $L_{\text{series}} = 5.9$ nH and $C_{\text{shunt}} = 1$ pF is introduced at each port of the network analyzer. The maximum measured isolation is 76 dB. The bandwidth at 40 dB isolation is measured as 0.4 MHz.

TABLE II
SUMMARY OF RESULTS AND COMPARISON TO PREVIOUS WORKS

| Metric | This work | [2] | [7] |
|---|---|---|---|
| Technology | FBAR | LC tank | CMOS |
| Operation frequency (MHz) | 2500 | 130 | 750 |
| Modulation frequency (MHz) | 3 | 40 | 750 |
| DC bias (V) | 0 | 1.7 | – |
| Modulation amplitude ($V_{pp}$) | 7 | 1.3 | 1.2 |
| Isolation (dB) | 76 | 65 | 50 |
| Insertion loss (dB) | 11 | 9 | 2 |
| Return loss (dB) | 8.5 | 4 | 12 |
| BW@40dB isolation (MHz) | 0.4 | 4 | 20 |
| Size ($mm^2$) | 96 | 374 | 25 |

The wye configuration's branch impedances necessitate the need for a matching network to optimize the circulator performance. Using a Smith chart, we calculate the matching network to be $L_{\text{series}} = 5.9$ nH and $C_{\text{shunt}} = 1$ pF. Fig. 7 shows the response of the circulator at port 1 after the matching network was programmed into the network analyzer. S21 has 11-dB insertion loss, while S31 achieves notch depth of 76 dB, demonstrating a nonreciprocity between ports 2 and 3 of 65 dB at port 1. The bandwidth at 40-dB isolation is 0.4 MHz.

## IV. CONCLUSION

In this letter, we demonstrate an FBAR circulator at 2.5 GHz which achieves 76-dB isolation with a 7-$V_{pp}$ modulation at 3 MHz. The insertion loss can be improved by a low loss PCB design using 3-D planar electromagnetic simulators such as Advanced Design System momentum and implementing a differential circulator topology [4], [5]. The FBARs in this circulator are zero-drift resonators with low $Q$ and $k_t^2$ due to an oxide compensation layer [6]. We can improve forward transmission and reverse isolation bandwidths using uncompensated FBARs ($Q_{\text{FBAR}} = 3000$ and $k_t^2 > 7\%$). Table II compares the FBAR circulator with first-of-type circulators in *LC* resonator and CMOS technologies. Modulating the FBAR using a varactor not only shifts the resonator frequency, but also its series resistance. We are currently analyzing if this effect can be harnessed to combine the advantages of both capacitive modulation and resistance modulation [7] circulator topologies. We also believe that nonreciprocity can be achieved using switches and capacitors in parallel instead of a varactor, simplifying the device further.

## ACKNOWLEDGMENT

The authors would like to thank Prof. D. Peroulis, Dr. M. A. Khater, and Dr. A. Fisher for their invaluable discussions on PCB design and experimental testing, and Prof. A. Alu and Prof. H. Krishnaswamy for sharing their insight and answering all our queries. They would also like to thank sspecially M. Quinn of Majelac Inc. for PCB assembly.